\documentclass[
  journal=largetwo,
  manuscript=article-type,
  year=2020,
  volume=37,
]{cup-journal}

\usepackage{amssymb}
\usepackage{amsmath}
\usepackage{gensymb}
\usepackage[nopatch]{microtype}
\usepackage{booktabs}
\usepackage[utf8]{inputenc}
\usepackage{rotating}
\usepackage{adjustbox}
\usepackage{afterpage}
\usepackage{subcaption}
\usepackage{stfloats}
\usepackage{multirow}
\usepackage{lineno}

\newcommand{\hi}{H{\sc i}}

\newcommand{\tspin}{$T_\text{spin}$}
\newcommand{\red}[1]{\textcolor{black}{#1}}

\newcommand{\nhi}{$N$\textsubscript{\sc{HI}}}

\newcommand{\kms}{km s$^{-1}$}

\newcommand{\tenpow}[1]{$\times 10^{#1}$}

\title{Detecting \hi\ Absorption in FRB Spectra: Modern Prospects and Scientific Utility}

\author{H. Roxburgh}
\affiliation{International Centre for Radio Astronomy Research, Curtin University, Bentley, WA 6102, Australia}
\email[H. Roxburgh]{hugh.roxburgh@postgrad.curtin.edu.au}

\author{M. Glowacki}
\affiliation{Institute for Astronomy, University of Edinburgh, Royal Observatory, Edinburgh, EH9 3HJ, United Kingdom}
\alsoaffiliation{International Centre for Radio Astronomy Research, Curtin University, Bentley, WA 6102, Australia}
\alsoaffiliation{Inter-University Institute for Data Intensive Astronomy, Department of Astronomy, University of Cape Town, Cape Town, South Africa}

\author{A. Bera}
\affiliation{International Centre for Radio Astronomy Research, Curtin University, Bentley, WA 6102, Australia}
\alsoaffiliation{ASTRON, the Netherlands Institute for Radio Astronomy, Oude Hoogeveensedijk 4,7991 PD Dwingeloo, The Netherlands}

\author{C. W. James}
\affiliation{International Centre for Radio Astronomy Research, Curtin University, Bentley, WA 6102, Australia}

\begin{document}

\begin{abstract}

Fast radio bursts (FRBs) emit broadband radio emission that may, in rare cases, encode atomic hydrogen (\hi) absorption signals as they traverse the interstellar medium of their host galaxies. \red{Though considered in the early FRB literature, the demanding observational prerequisites and the rarity of suitable events have meant that no thorough search for \hi\ absorption in FRB spectra has yet been undertaken. Here, we present an updated systematic analysis assessing the likelihood of modern facilities to detect such absorption features. As a proof of concept, we search for absorption in the spectrum of the bright ASKAP-localised FRB~20211127I, finding a $3\sigma$
opacity upper limit of 0.51. While this test case offers little constraining power, we find that narrow FRBs with fluences exceeding 20/70/150 Jy ms observed with MeerKAT/ASKAP/DSA can probe opacities below 0.1 --- a regime in which absorption detections become physically meaningful. We further highlight that stacking thousands of bursts from hyperactive repeaters with FAST offers a very powerful avenue toward detection. Finally, we discuss the broad scientific potential of such detections, including constraints on extragalactic \hi\ spin temperatures, a means to physically probe the environment surrounding the progenitor, and a path towards disentangling host galaxy contributions to dispersion and scattering.}

\end{abstract}

\section{Introduction}\label{sec:intro}

\red{Fast radio bursts \citep{Lorimer2007,Thornton2013} have emerged as powerful probes of baryonic matter in the Universe, owing to their dispersion measures (DMs) which directly encode the integrated free electron density along the line of sight. Acting as backlights, they illuminate the diffuse gas occupying the halos and outskirts of foreground galaxies, material that is otherwise extraordinarily difficult to observe \citep{Macquart2020,Khrykin2024,Connor2025}.}

\red{However, it is also possible that contained within their spectra lies valuable information concerning the conditions of the neutral material in the ISM of their host galaxies. Such information would manifest through an atomic hydrogen (\hi) absorption line, which arises from the spin-flip transition at a rest wavelength of 21 cm. Neutral hydrogen constitutes the dominant baryonic component of the ISM and serves as the raw material for star formation \citep{Krumholz09}; its properties and dynamics are therefore intrinsically tied to the evolution of galaxies across cosmic time. The depth and width of absorption features directly constrain the thermal and physical conditions of this gas \citep{Allison2021}, and thus encode information that offers direct insights into galaxy evolution. Furthermore, absorption probes much finer scales than 21-cm emission observations, which reveal only the bulk distribution and kinematics of \hi\ gas and are highly limited by sensitivity at cosmological distances.} 

\red{As FRBs are detected across a wide range of redshifts \citep[out to $z\sim2$,][]{Ryder2023,Caleb2025} and host galaxy types \citep{Gordon2023,Sharma2024}, they represent a potentially powerful and diverse population of background sources through which \hi\ absorption can be studied. Furthermore, they offer inherently unique benefits compared with typical absorption backlights such as active galactic nuclei (AGN). For example, their transient nature trivialises the often tricky process of disentangling \hi\ absorption and emission profiles, and their extremely compact sizes \citep[$r\sim10$ km,][]{Farah2018} make them true point sources, thereby removing the uncertainty of covering factor \citep{Allison2021} when interpreting the physical ramifications of a measured optical depth.} 

\red{The prospect of \hi\ absorption in FRBs was discussed by \citet{Fender2015,Margalit2016} more than a decade ago, during the very early years of the FRB field. At this time, no burst had yet been localised to a host galaxy, and the existence of a repeating class of FRBs was only in the early process of establishment \citep{Spitler2016}. Consequently, both works focused primarily on the prospect of using \hi\ absorption itself to constrain host galaxy redshifts, a motivation now entirely superseded by the rapid localisation capabilities of contemporary observatories \citep{Bannister2019,Kocz2019}. As such, searches for \hi\ absorption in FRBs have been largely overlooked; to date, only one burst has been reported to have had its spectrum investigated \citep{Osowski2019}. In this case, as the host redshift was unknown, the search spanned the entire observing band without targeting a specific frequency, and no firm evidence of absorption was identified.}

\red{Even targeted searches for host-galaxy-associated \hi\ absorption face a restrictive set of circumstances that must simultaneously be met. First, as the bandwidth an absorption line would occupy ($\lesssim200$ kHz) falls below the coarse channel width FRBs are typically detected with ($\sim$1 MHz), voltage capture is required to enable coherent post-processing and the formation of high-resolution spectra. Second, the captured voltages must support arcsecond-scale localisation of the burst, confirming the host galaxy and thus the redshifted frequency of the \hi\ line. Third, that frequency must fall within the observing band of the dataset in which the FRB is detected, and --- given their diverse spectral structure \citep{Pleunis2021} and the interstellar scintillation they may exhibit \citep{Masui2015} --- must coincide with a region of locally high SNR within the band. Additional constraints arise from the physical nature of the host ISM and the FRB itself; any detection naturally requires the presence of a sufficiently dense \hi\ medium along the line of sight, as well as a burst bright enough to probe the shallow optical depths typical of such environments.}

\red{Nevertheless, as the number of discovered FRBs continues to grow rapidly \citep{Shannon2024,Connor2025,Pastor-Marazuela2025,CHIME-dr2}, so too does the probability of suitable events occurring. In parallel, the scope of FRB science has expanded dramatically, driven in large part by their utility as unbiased probes of cosmic structure \citep{Macquart2020,James2022}. \hi\ absorption may offer unique insight into several of its emerging frontiers, including the local environments of FRB progenitors and the disentangling of host galaxy DM contributions from the cosmic DM for improved cosmological inference. It is therefore timely to revisit the concept under a modern lens.}

\red{As such, in this paper we explore the prospect of detecting \hi\ absorption in FRBs and outline the potential scientific promise of such a detection. In Section \ref{sec:theory}, we reintroduce the theory behind measuring \hi\ absorption in FRB spectra with respect to commonly used parameters in the field.  In Section \ref{sec:data}, we present a proof of concept using a bright localised FRB that satisfies all the required criteria. In Section \ref{sec:discussion}, we investigate the outlook for observing \hi\ absorption in the current and future eras, discussing detection probabilities across facilities and the particular promise of repeating FRBs. In Section \ref{sec:science}, we highlight the scientific value of such detections, spanning spin temperature constraints, insights into progenitor environments, and the prospect of gauging an FRB's depth within its host galaxy. Finally, we conclude in Section \ref{sec:conclusion}.}

\section{Theory}\label{sec:theory}

When an FRB with a variable spectral flux density $S_\text{FRB}$ passes through a foreground \hi\ gas reservoir, the strength of absorption occurring as a function of frequency is quantified by the optical depth

\begin{equation}
    \tau(\nu) = -\ln\Biggl(1-\frac{\Delta S(\nu)}{f_c~S_\text{FRB}(\nu)}\Biggr)\ ,
\end{equation}

\noindent where $\Delta S$ is the difference between $S_\text{FRB}$ and the minimum of the absorption line, and $f_c$ is the covering fraction --- the fraction of the background source covered by the foreground absorber on the sky. As FRBs are true point sources \citep{Farah2018}, $f_c$ is always equal to unity, so we drop it henceforth. For optically thin emission ($\tau<<1$), this simplifies to 

\begin{equation}\label{eq:simple}
    \tau(\nu) \simeq \frac{\Delta S(\nu)}{S_\text{FRB}(\nu)}\ .
\end{equation}

A telescope's ability to detect an absorption feature is solely dependent on the SNR it observes an FRB with. Specifically, the SNR of interest is that of the FRB's spectrum averaged over its temporal pulse width $w_\text{FRB}$ at a spectral resolution of $\Delta V$ in velocity space. Defining $\sigma_\nu$ as the 1$\sigma$ uncertainty on the FRB flux density in each frequency channel, and $\sigma_\tau$ as the corresponding 1$\sigma$ uncertainty on the derived optical depth, Eq.~\ref{eq:simple} can then be written as

\begin{equation}
    \sigma_\tau = \frac{\sigma_\nu}{S_\text{FRB,HI}}= \frac{1}{\text{SNR}_\text{FRB,HI}}\ .
\end{equation}

\noindent Here, the subscript emphasises that this is the spectral SNR measured at the location of the \hi\ line. This is not the same SNR as is reported in FRB discovery papers, which only \red{relates to the integral over the temporal pulse profile and not to its continuum spectrum}. Therefore, while we stress that SNR is the sole deciding factor in an FRB's sensitivity to absorption, if we wish to estimate the sensitivities probed by the current population of detected FRBs, we must expand this expression in terms of their reported properties and those of the theoretical absorption features.

The noise in the pulse-averaged spectrum can be derived from the 
radiometer equation:

\[
    \sigma_\nu = \frac{\text{SEFD}}{\sqrt{N_\text{pol}\ w_\text{FRB}\ 
    \Delta\nu}}\ \ \text{Jy}
\]

\begin{equation}
    = \text{SEFD} \sqrt{\frac{c}{\nu_\text{HI}}}\sqrt{\frac{1+z_\text{FRB}}
    {N_\text{pol}\ w_\text{FRB}\ \Delta V}}\ \ \text{Jy} \ ,
\end{equation}

\noindent after transforming into the FRB's rest frame velocity space 
using its redshift $z_\text{FRB}$. 

\red{While the true strength of absorption is quantified by the integration of optical depth over the full line width $W$, for clarity we choose to remain in the dimensionless optical depth space and quantify absorption strength through the line-averaged opacity $\bar{\tau}$, whose uncertainty is lower than $\sigma_\tau$ by a factor of $\sqrt{N}=\sqrt{W/\Delta V}$, where $N$ denotes the number of a channels spanned by the absorption line.} Thus, we arrive at an expression for the sensitivity to absorption opacity in terms of a telescope’s SEFD, the line width, and an FRB’s properties:

\[
    \sigma_{\bar{\tau}} = \frac{\sigma_\nu}{S_\text{FRB,HI}}\sqrt{\frac{\Delta V}{W}}
\]
\begin{equation}\label{eq:penultimate}
    = \frac{\text{SEFD}}{S_\text{FRB,HI}}\sqrt{\frac{c}{\nu_\text{HI}}}\sqrt{\frac{1+z_\text{FRB}}{N_\text{pol}\ w_\text{FRB}~W}} \  .
\end{equation}

The final simplification we can incorporate deals with the spectral behaviour of FRBs, which can be complex; some bursts occupy only part of the observing bandwidth \citep{Pleunis2021}, and others exhibit scintillation which \red{redistributes flux across the band, and may either suppress or enhance the flux density at the \hi\ line frequency} \citep{Masui2015}. However, while there is some evidence for a preference toward lower-frequency emission \citep{James2022,Shin2023,Shannon2024}, no strong systematic trend in flux density is observed across the narrow bandwidth relevant here. For simplicity, and following several previous studies \citep{CHIMEDR1}, we therefore can assume an idealised FRB with a flat spectrum, such that $S_\text{FRB,HI}(\nu)=\bar{S}_\text{FRB,HI}(\nu)$. This can be expressed in terms of the FRB fluence, $F_\text{FRB}$, as
\[
\bar{S}_\text{FRB,HI}(\nu) = \frac{F_\text{FRB}}{w_\text{FRB}}\ ,
\]

\noindent which allows Equation~\ref{eq:penultimate} to be recast in terms of the observed FRB fluence. In this form, the corresponding 3$\sigma$ limit becomes

\begin{equation}\label{eq:final}
    L_{3\sigma} = \frac{3\times\text{SEFD}}{F_\text{FRB}}\sqrt{\frac{c}{\nu_\text{HI}}}\sqrt{\frac{ w_\text{FRB}\ (1+z_\text{FRB})}{N_\text{pol}~W}} \ .
\end{equation}

\noindent Expressing the sensitivity in this form is useful because FRB fluences are routinely reported in the literature, and thus such a metric enables straightforward estimates of \hi\ absorption sensitivity for existing FRB samples.

\section{FRB~20211127I: A Test Case}\label{sec:data}

\red{We now conduct a preliminary search for \hi\ absorption in the spectrum of a real FRB observed with the Australian SKA Pathfinder \citep[ASKAP;][]{ASKAP} telescope. To select a suitable candidate, we again consider the mandatory requirements for the detection of an associated\footnote{FRBs whose redshift is unknown and whose full spectrum must thus be searched \citep{Osowski2019} are not considered here.} absorption feature in an FRB's spectrum:} 

\begin{enumerate}
    \item The spectrum must be captured \red{at a resolution comparable to or finer than a typical absorption line width ($\Delta V\sim50$ \kms, or $\Delta\nu \sim 200$ kHz at $z=0.1$) to avoid significant sensitivity loss from spectral dilution.}
    \item The localisation must be precisely understood (i.e. typically within a few arcseconds) to determine the redshift of the host galaxy and thus the redshifted frequency of the \hi\ line.
    \item There must be non-negligible FRB signal at that redshifted frequency within the observed bandwidth.
\end{enumerate}

\red{For this proof of concept, we focus on the CRAFT-ICS sample of 43 FRBs observed between 2018 and 2024 \citep{Shannon2024}, 37 of which had their raw voltages captured. Of these, just 5 meet the three criteria: FRB~20190608B, FRB~20210117A, FRB~20211127I, FRB~20230526A, and FRB 20230718A. As outlined in Eq.~\ref{eq:final}, lower opacity limits (i.e. improved absorption sensitivities) scale with $Fw^{-1/2}$; the values for these five bursts are 8.5, 19.0, 50.4, 20.7, and 16.8 Jy ms$^{-1/2}$ respectively\footnote{We use widths from \citet{Scott2025}, which presents the more accurate high time resolution properties of the CRAFT-ICS sample.}, making FRB~20211127I the clear best candidate for our investigation.} 

\begin{figure*}[!t]
    \centering
    \includegraphics[width=\linewidth]{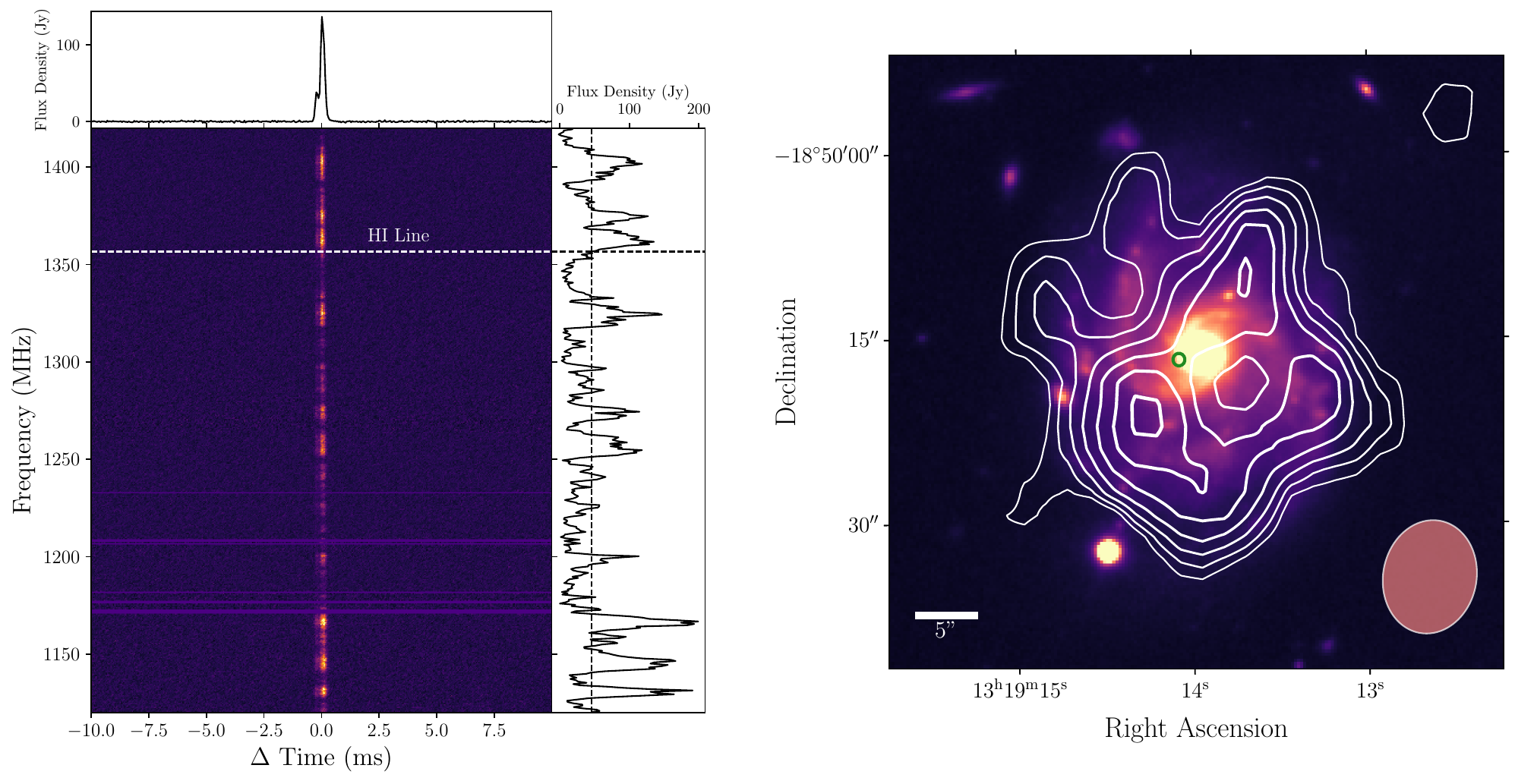}
    \caption{(Left) Dynamic spectrum of FRB 20211127I. The right panel displays the spectrum of the FRB averaged across its temporal pulse width. The location of the redshifted \hi\ line is shown by the horizontal dashed line, and the band-averaged flux density is denoted by the vertical dashed line in the right panel. (Right) VLT $i$-band image of the host galaxy of FRB 20211127I overlaid with its \hi\ emission as seen by MeerKAT (white contours) and the localisation region in green. Contours increase as [1,1.25,1.5,...] $\times$ 5.34 \tenpow{20} cm$^{-2}$ and the beam is shown in the bottom right.}
    \label{fig:combinedfig}
\end{figure*}

\red{This FRB is also among the best-studied bursts in the sample, with several works having examined its spectral morphology and host galaxy properties \citep{Glowacki2023,Shannon2024,Scott2025,Roxburgh2025}. Notably, a substantial \hi\ reservoir has been confirmed in the host via emission mapping, further motivating this FRB as a target for an absorption search.} In Figure \ref{fig:combinedfig}, we present its dedispersed dynamic spectrum, which has been processed offline through the CELEBI \citep[CRAFT Effortless Localisation and Enhanced Burst Inspection;][]{Scott2023,Glowacki2026} pipeline. Alongside this, we display VLT imaging of its host galaxy with its localisation and the host's \hi\ emission overlaid as measured by MeerKAT \citep{MeerKAT} with a 3 hr $L$-band observation.

Unfortunately, as is evident in the ``striped" appearance of the FRB's signal in the dynamic spectrum, this FRB exhibits strong interstellar scintillation. Some FRBs show scintillation patterns with decorrelation bandwidths $\nu_\text{DC}$ --- i.e. the scale over which scintillation fringes oscillate --- near or equal to the characteristic widths of absorption features. \citet{Scott2025} find a $\nu_\text{DC}$ of 2.9 MHz for this FRB, which corresponds to around 650 \kms\ in the rest frame, and thus this is not a significant issue for this FRB. However, scintillation still makes discerning an absorption signal difficult; as shown by the horizontal line in the dynamic spectrum, the \hi\ line lies near the boundary of a scintle and a trough. Regardless, we can extract a high resolution spectrum around this region and search for absorption signals. 

\begin{figure}[!t]
    \centering
    \includegraphics[width=\linewidth]{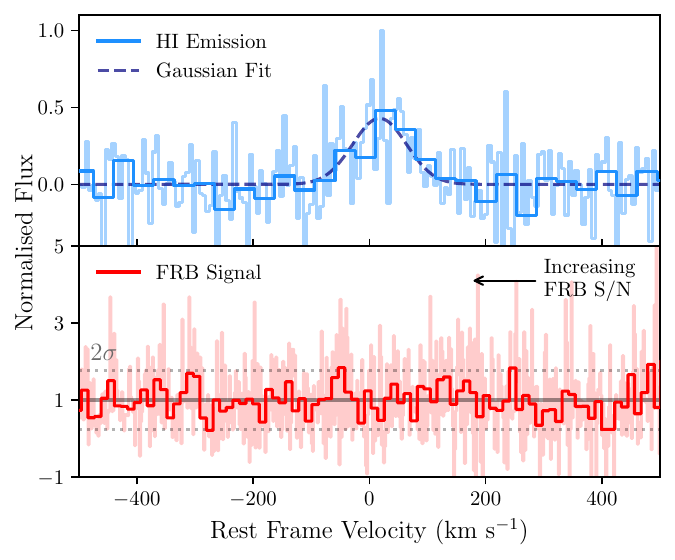}
    \caption{(Top) Normalised \hi\ emission measured by MeerKAT at the pixel closest to the localisation region. The raw data is overlaid with a Gaussian profile fit to the binned data. (Bottom) Flattened and normalised FRB~20211127I spectrum at the location of the \hi\ line. For visual purposes, the spectrum is overlaid with a binned spectrum at a velocity resolution of 11 \kms\ in the rest frame and the 2$\sigma$ deviation from unity is shown. Due to scintillation, the SNR of the spectrum increases towards the left.}
    \label{fig:spectra}
\end{figure}

In Figure \ref{fig:spectra}, we present the \hi\ emission line spectrum arising from the pixel nearest to the FRB's localisation region. We also show the same frequency range in the FRB's dedispersed signal, which has been flattened \red{to remove the broad scintillation variation} by fitting a local \red{2nd order} polynomial baseline, calculated while excluding all emission channels. As shown, we find no evidence for an \hi\ absorption feature in the spectrum of FRB 20211127I. 

Instead, we can estimate an upper limit \red{on the strength of \hi\ absorption by directly measuring $\sigma_{\bar{\tau}}$}. If we assume an \hi\ line width $W$ of 50 \kms\ --- which is broadly a representative width of \hi\ absorption features detected in spectra of extragalactic sources \citep{Prochaska2008} --- we can rebin the normalised absorption spectrum such that our individual channel width is equal to $W$ and measure the resulting noise level. \red{We find this to be 0.17, and therefore our 3$\sigma$ opacity upper limit is equal to 0.51.}

\red{This limit is quite unremarkable, and thus we choose not to thoroughly search the remaining weaker CRAFT-ICS FRBs, saving such investigations for future studies with larger samples where spectral stacking across bursts becomes viable. In the following section, we comment on the specific pros and cons of this particular FRB, and the outlook of detecting \hi\ absorption and constraining \tspin\ with future FRB observation campaigns.}

\section{The Prospect of \hi\ Absorption in FRBs}\label{sec:discussion}

\subsection{Strengths and Limitations of FRB 20211127I}

FRB 20211127I is in many ways an ideal burst for such an investigation. It is a bright and narrow burst, exhibiting little to no scattering, which allows for a high SNR in the pulse-averaged spectrum. Even the dominant interstellar scintillation pattern noted earlier does not present a major barrier in this case. Scintillation typically complicates absorption searches, and this FRB exhibits one of the strongest patterns in the CRAFT sample; \citet{Scott2025} report a spectral modulation index of $m = 0.74$ (values near unity indicate strong scintillation). Even so, its impact on the effective sensitivity of this test case is minimal. As shown by the FRB spectrum in the left image of Figure ~\ref{fig:combinedfig}, the flux density of the spectrum at the location of the \hi\ line almost perfectly coincides with the band-averaged flux density of the FRB. As such, our 3$\sigma$ limit would not change significantly if the scintillation was weaker or even non-existant. In fact, scintillation of this manner could, in a different scenario, be considered an advantage rather than a disadvantage; for FRBs with $\nu_\text{DC}$ values significantly larger than the typical width of absorption features, as is the case for FRB 20211127I, a coincidental alignment of a scintle with the \hi\ line could provide a boost in SNR across the line that could significantly alter the likelihood of detection.

To illustrate this, we consider the resulting limits on both opacity and \tspin\ were the \hi\ line to coincide with the local spectral maxima near 1362 MHz, rather than at 1357 MHz. In this case, the SNR increases by roughly a factor of 2, yielding an opacity limit of approximately 0.25 and a corresponding \tspin\ lower limit of 60 K. Such an opacity is still very much on the extreme boundary of plausible detection \citep{Braun2012}, but the resulting \tspin\ limit would be within $\sim$15 K of temperatures observed in the Milky Way’s coldest atomic clouds \citep{Murray2018}. 

\subsection{Detectability of HI Absorption in Single FRBs}\label{subsec:non-repeaters}

\red{Here we focus our discussion on the feasibility of detecting \hi\ absorption in the spectrum of any given single burst, either a non-repeating FRB or a chosen repeat burst from a repeating FRB. We will discuss the implications of the repeating FRB population later in Section \ref{subsec:non-repeaters}.}

\subsubsection{Current Facility Capabilities}\label{subsec:current}

As outlined in Section \ref{sec:theory}, the detectability of \hi\ absorption in FRB signals is governed by a combination of instrumental and source properties, including the system noise, the \hi\ line width, and the FRB’s intensity, temporal duration, and redshift. To understand the broader likelihood of any given facility to detect absorption, we must also consider their fields of view (FoV) and frequency ranges, as well as the intrinsic luminosity distribution and spectral behaviour of the FRB population.

Although absorption measurements are not limited by the distance-dependent sensitivity constraints that restrict \hi\ emission studies to the local Universe, \red{it remains essential to identify the host galaxy in order to constrain the redshift and associate any inferred gas properties with a well-understood sightline. It is therefore important to consider both the observing bands of each telescope and the likelihood of localising an FRB whose host-galaxy redshift falls within those bands. Consequently, for the remainder of this analysis, we exclude data from the Canadian Hydrogen Intensity Mapping Experiment \citep[CHIME;][]{CHIME}, which observes between 400-800 MHz. Despite detecting the majority of FRBs, CHIME is exceedingly unlikely to localise bursts at sufficiently high redshift ($z>0.78$) for the \hi\ line to fall within this observing band.} However, we note that its successor, The Canadian Hydrogen Observatory and Radio-transient Detector \citep[CHORD;][]{CHORD}, will operate with an ultra-wide band between 300-1500 MHz and thus is poised to become an important resource for future work. Henceforth, we consider only the current capabilities of the remaining major facilities in the field --- ASKAP, the Deep Synoptic Array \citep[DSA;][]{Kocz2019}, the Five-hundred-meter Aperture Spherical Telescope \citep[FAST;][]{FAST}, and MeerKAT. \red{Each of these telescopes are capable of observing the \hi\ line over a broad range of redshifts, including the local Universe between 1.3-1.4 GHz.}

Our aim is to estimate the likelihood that a single FRB detected by each of these facilities will be bright enough to probe realistic \hi\ absorption. By calculating this probability, we can quantify the broader relative ability of these different observatories to conduct such studies. Eq.~\ref{eq:final} provides us with the tools to make such estimates; all we require are the SEFDs of the telescopes. Since ASKAP, DSA, and MeerKAT each have the capability to process FRBs offline after detection, we can assume the full sensitivity achievable through coherent beamforming. At current maximum operating capacities (36, 63, and 64 antennas, respectively), their effective SEFDs are approximately 50~Jy, 100~Jy, and 12.5~Jy, while FAST achieves around 1.25~Jy at $L$-band. 

\begin{figure*}[t]
    \centering
    \includegraphics[width=\linewidth]{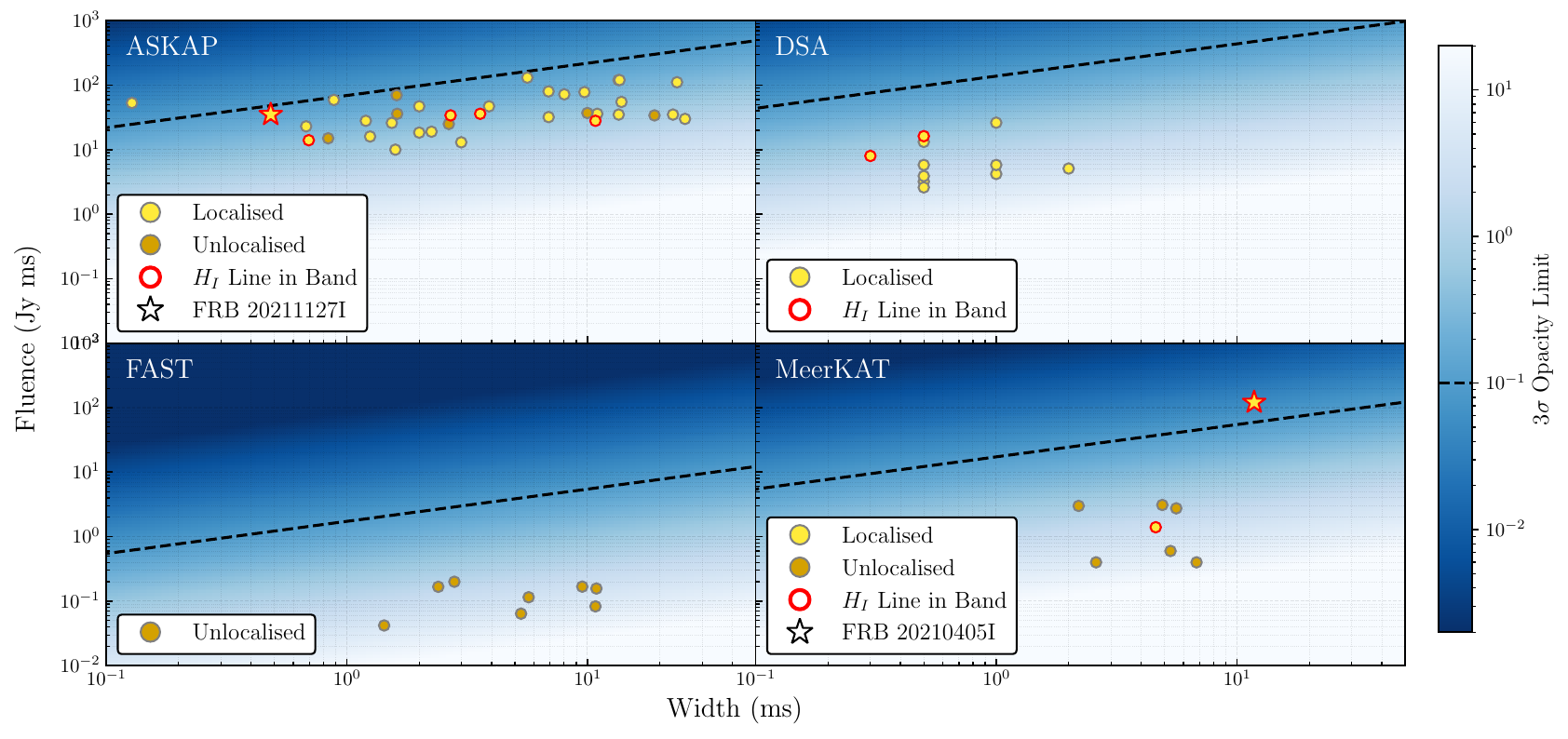}
    \caption{3$\sigma$ limits on the \hi\ \red{opacity} detectable in the pulse-averaged spectra of single FRBs observed by various telescopes. The black dashed lines in each panel indicate an \red{opacity limit of 0.1}. These limits assume maximal sensitivity (i.e. with full antenna configuration) and coherent beamforming, an \hi\ line width of 50 \kms, and a flat FRB spectrum. Overlaid are samples of reported FRBs with known fluences and widths: ASKAP \citep{Shannon2024, Scott2025}, DSA \citep{Law2024}, FAST \citep{Zhu2020,Niu2021,Zhou2023}, and MeerKAT \citep{Rajwade2022,Jankowski2023,Driessen2024}. \red{FRB markers are coloured by localisation status; red outlines denote events where the host galaxy redshift places the \hi\ line within the FRB's observed bandwidth.}}
    \label{fig:limits}
\end{figure*}

Once again setting the assumed absorption line width $W$ to 50 \kms, Figure \ref{fig:limits} presents grids of 3$\sigma$ \red{opacity} limits as a function of burst width and fluence for each telescope at maximum observing sensitivity, calculated using Eq.~\ref{eq:final}. We overlay FRBs with reported widths and fluences to contextualise current samples, noting that the recent sample expansions for DSA \citep{Sharma2024,Connor2025} and MeerKAT \citep{Pastor-Marazuela2025} do not currently include such information. 

\red{The background shading represents the derived opacity limits, with darker blue regions (toward the upper-left) corresponding to more stringent constraints that probe increasingly faint absorption features. The dashed black line in each panel marks a $3\sigma$ opacity limit of 0.1, which we consider a threshold for plausible absorption detection. Localised and unlocalised bursts are separated to identify FRBs whose known host redshift places the \hi\ line within the observed bandwidth --- these are highlighted with red borders. We do note that because some FRBs were observed with a reduced antenna count, their true opacity limits are larger (less constraining) than the background suggests; consequently, these events appear artificially high relative to the underlying colouration.}

\red{While few known FRBs currently lie in the regime where useful opacity limits can be placed, many approach this threshold, suggesting that the technique will become increasingly viable as FRB samples grow.} In fact, one event stands out clearly in this figure: the MeerKAT FRB~20210405I. This burst was exceptionally bright, and, having been localised to redshift 0.06 galaxy \citep{Driessen2024}, contains the \hi\ line frequency within its bandwidth. Furthermore, \citet{Roxburgh2025} conducted follow-up observations on the host with MeerKAT to reveal a clear \hi\ disk, confirming the presence of a sizeable \hi\ reservoir. Thus, this FRB would be an excellent candidate for probing \hi\ absorption and for potentially constraining the host galaxy \tspin. Unfortunately, the burst was detected prior to the implementation of MeerTRAP’s transient buffer system \citep{Rajwade2024}, which now enables the recording of full-voltage data for later processing. Consequently, only a coarse 1024-channel spectrum spanning $\sim800$~MHz is available, providing insufficient spectral resolution for a meaningful absorption search. Nonetheless, this event demonstrates that such favourable FRBs do occur, and that similar sources will be accessible to future absorption studies with the voltage-capture capabilities now available on ASKAP, DSA, and MeerKAT.

To calculate the fraction of detected FRBs capable of probing HI absorption, we must define an SNR threshold, $\text{SNR}_\text{thr}(w)$, required to surpass a target $3\sigma$ opacity limit, $\bar{\tau}_\text{thr}$. We then can estimate the probability $P(\text{SNR} \ge \text{SNR}_\text{thr}(w) \mid \text{SNR} \ge \text{SNR}_\text{det})$, where $\text{SNR}_\text{det}$ is the detection threshold of the facility. This estimation relies on the SNR distributions observed by each telescope, which reflect the underlying FRB fluence distribution. If this distribution is scale-invariant above each telescope's fluence threshold ($F_\text{det}$), all facilities observe SNR distributions of identical shape. Consequently, the fraction of absorption-sensitive bursts scales as $(F_\text{thr}(w)/F_\text{det})^{\gamma+1}$, where $\gamma$ is the differential power-law index.

Several studies have attempted to model the fluence distribution of the FRB population \citep[e.g.][]{James2019,Shin2023,Ryder2023,Arcus2025}, generally employing a Schechter function to determine the high end cutoff energy imposed by the expected physical limit to the FRB emission mechanism. This function is not scale invariant, but given the downturn must lie orders of magnitude above telescope thresholds \citep{Arcus2025}, we can consider the SNR distributions effectively equivalent across instruments. Possible deviations at low fluences --- such as variations in the power-law slope reported for some repeaters \citep{Li2021,Kirsten2024} and for non-repeating FRBs detected by MeerKAT \citep{Jankowski2023} --- remain inconclusive and are neglected here. Studies in the $L$-band generally find $\gamma$ values near -2 \citep{James2022,Hoffmann24,Arcus2025}, as expected for a cosmological population of sources. Thus, the fraction of interest reduces simply to $F_\text{det}/F_\text{thr}(w)$.

As $F_\text{thr}$ is a function of $w$, we must multiply this fraction by the probability distribution of pulse widths $p(w)$ and then integrate over $w$. Ideally we would use each telescope's observed width distribution; due to the current low number of reported widths in DSA, FAST, and MeerKAT, we fit a log normal function to the width distribution of ASKAP and assume this for all. Now only the fluence detection thresholds of each telescope and observing system are required \citep{Shannon2024,Wang2025,Connor2025,Niu2021,Jankowski2023}. As only very bright FRBs can probe low $\bar{\tau}_\mathrm{thr}~$, such events are detectable even with the incoherent summation modes used by ASKAP and MeerKAT; in these modes, $F_\text{det}$ is a factor of $\sqrt{N_\text{ant}}$ closer to $F_\text{thr}$, which doesn't change because it is always defined with respect to the telescopes' coherent sensitivity due to the fact that the array facilities can process offline after detection. 

Table \ref{tab:fom} presents the resulting absorption-probing fractions of the four observatories in their different detection modes as percentages ($\bar{\tau}~\%$), given a target \red{$\bar{\tau}_\mathrm{thr}$ of 0.1}. From this we can see that a sizeable fraction of FRBs detected with the incoherent modes of ASKAP and MeerKAT are expected to probe realistic \hi\ absorption. Of course, there are the extra factors concerning the redshift distribution and observation band that further reduce the fraction of usable bursts; however, these numbers indicate that absorption detection in the spectra of FRBs is certainly not implausible.

To compare the relative likelihood of detection between facilities, we must simply estimate the total number of FRBs with fluences greater than $F_\text{thr}$ each facility will detect in an equal observing time. This figure of merit (FoM) is effectively the total number of FRBs a system detects multiplied by the absorption-probing fractions calculated above. The former is proportional simply to the FoV / $F_\text{det}~$, and thus the FoM is given by

\begin{equation}
    \label{eq:fom}
    \mathrm{FoM}
    = \int \frac{\mathrm{FoV}}{F_\mathrm{thr}( w)}
    \, p(w) \, \mathrm{d}w \ .
\end{equation}

\noindent The resulting values, also found in Table \ref{tab:fom}, illustrates that ASKAP, observing in either incoherent or coherent mode, is by far the most likely telescope to probe \hi\ absorption in the spectra of single FRBs. \red{Furthermore, the results stress the importance of FoV for this analysis; even a modest increase in surveyed solid angle can outweigh a significant gain in per-burst sensitivity, since the opacity limit degrades only as $\sqrt{\text{SEFD}}$ while the detection rate scales linearly with FoV. This highlights that wide-field instruments, which are more feasible using incoherent detection modes, represent a valuable design choice for this science case.}

\begin{table}[t]
\begin{threeparttable}
\caption{Comparison of facility-specific sensitivity to \hi\ absorption. The fourth column indicates the percentage of detected FRBs with fluences great enough to probe an opacity threshold ($\bar{\tau}_\text{thr}$) of 0.1. The Figure of Merit in the final column indicates the relative likelihood of detecting an FRB that probes the same $\bar{\tau}_\text{thr}$ in an equal observing time, calculated using Eq. \ref{eq:fom}.}
\label{tab:fom}
\begin{tabular}{l|ccccc}
\toprule
\headrow Facility &  Detection Mode & $F_\text{det}$ (Jy ms)\tnote{a} & $\bar{\tau}~\%$  &FoV (deg$^2$)\tnote{a} & FoM \\
\midrule
& Incoherent & 8.0 & 6.2 & 30  & 23.2\\
\multirow{-2}{*}{ASKAP} & Coherent & 1.2 & 0.9 & 30 & 23.2 \\
\midrule
DSA\tnote{$\dagger$} & Coherent  & 1.9 & 0.8 & 3.4 & 1.32 \\
\midrule
FAST    & Single dish   & 0.015  & 0.5     & 0.008 & 0.25 \\
\midrule
  & Incoherent & 3.4 & 10.6 & 1.3 & 4.02 \\
\multirow{-2}{*}{MeerKAT} &  Coherent & 0.7 & 2.0 & 0.4 & 1.23 \\
\midrule
\end{tabular}
\begin{tablenotes}
    \item[a] Values are taken from \citet{Shannon2024,Wang2025,Law2024,Connor2025,Niu2021,Jankowski2023}. 
    \item[$\dagger$] DSA only observes in coherent detection mode. 
\end{tablenotes}
\end{threeparttable}
\end{table}

\subsubsection{Future Facility Capabilities}

Estimating the performance of future facilities like SKA-Mid or the full DSA is challenging \red{and likely impractical}, as it depends heavily on the specific configurations of their FRB detection backends, \red{which are currently unspecified}. However, the limits presented in Figure \ref{fig:limits} are independent of detection architecture; provided the coherent SEFD of a future facility can be estimated, similar sensitivity projections can be made. 

The predicted SEFDs for SKA-Mid\footnote{\url{https://www.skao.int/en/resources/technical-documents}} and the 1650-dish DSA\footnote{\url{https://www.deepsynoptic.org/}} are approximately 2.5 Jy and 4 Jy, respectively. At these sensitivity levels, a 1 ms burst would require a fluence of 5 Jy ms (SKA-Mid) or 8 Jy ms (DSA) to probe a 50 \kms\ wide absorption line at a 3$\sigma$ opacity limit of 0.1. 


\subsection{Repeating FRBs}

The population of repeating FRBs poses an entirely different opportunity for probing \hi\ absorption. Dozens of these sources have now been detected \citep{Spitler2016,Repeaters,Cook2026}, with some "hyperactive" repeaters \citep{Konijn24,Tian2025} exhibiting thousands of outbursts. As these bursts originate from the same environment and traverse identical structures within the host galaxy, stacking their signals increases the cumulative integration time for any imprinted absorption features, effectively boosting the sensitivity to the \hi\ line. This clearly makes hyperactive repeaters --- several of which have been localised to host galaxies with redshifts below 0.1 \citep{Day2021,Ravi2023} --- the best targets for \hi\ absorption. 

\begin{figure}[t]
    \centering
    \includegraphics[width=\linewidth]{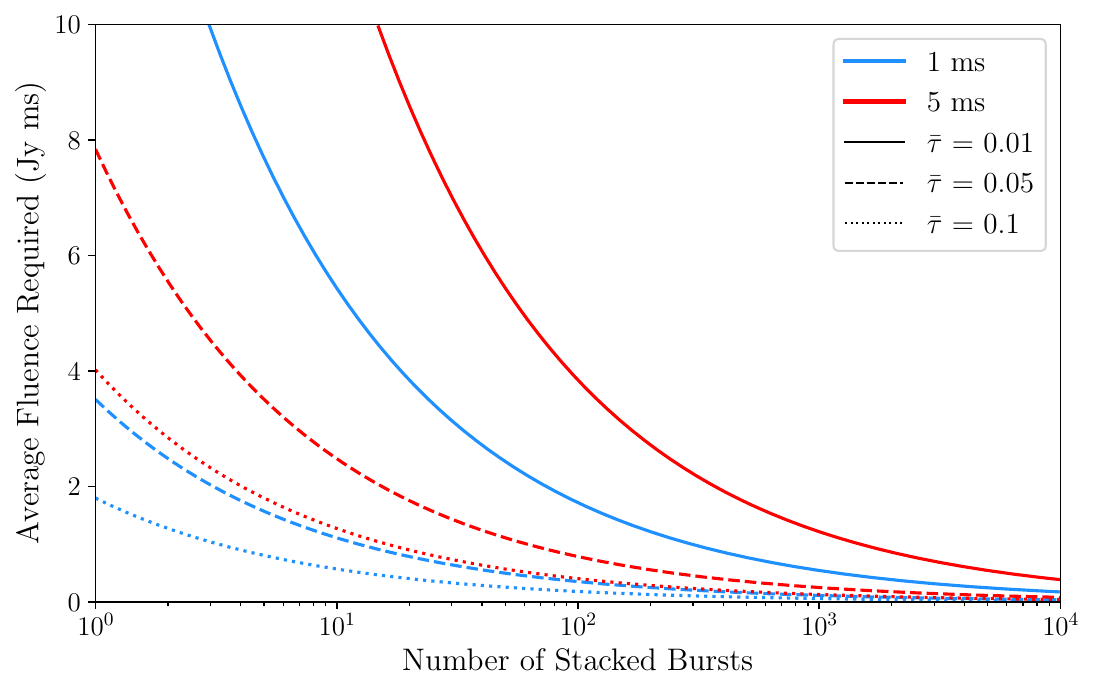}
    \caption{Average burst fluence required for FAST to detect various integrated \hi\ optical depths as a function of the number of bursts stacked from a repeating FRB.}
    \label{fig:stacked}
\end{figure}

With its incredible sensitivity, FAST's true strengths are not in the serendipitous detection of apparently non-repeating FRBs, but in the regular monitoring of these repeating FRBs. If we assume that each burst has the same pulse width \citep[for simplicity; width can vary by an order of magnitude, e.g.][]{Li2021}, then the improvement in sensitivity to \hi\ absorption scales with $\sqrt{N_\text{bursts}}$. As such, we can estimate how FAST's fluence limit for detecting a given opacity changes as a function of the number of stacked FRBs, which we demonstrate in Figure \ref{fig:stacked}. This shows that just 100 FRBs with a pulse width of 1 ms and an average fluence of 2 Jy ms can probe opacities of just 0.01. Of course, we reiterate that this estimate uses idealised, flat spectrum FRBs.

Regardless, there already exists known ideal candidates that are highly sensitive probes of absorption. For example, the $z=0.077$ repeater FRB 20220912A has outbursted thousands of times with fluences around 1 Jy ms as observed by FAST \citep{Zhang2023}, and several tens of times above 50 Jy ms as observed by the Allen Telescope Array \citep{Sheikh2024}. \red{While it has exhibited highly varied burst morphology, given its redshifted \hi\ frequency falls at around 1320 MHz --- just 70 MHz from the centre of FAST's $L$-band --- this target would be perfect for both \hi\ absorption and \tspin\ investigations with FAST.}

Furthermore, with CHIME's recent second data release reporting dozens of new repeaters \citep{CHIME-dr2,Cook2026}, the sample of hyperactive repeaters is bound to continue to expand, making absorption detection increasingly realistic.

\section{Potential Science with FRB \hi\ Absorption}\label{sec:science}

\citet{Fender2015,Margalit2016} previously identified the use of potential \hi\ absorption in FRBs as an independent distance estimator. The rapid improvements in localisation ability since their release have reduced the value for such a concept, \red{though very distant sources with no identifiable host galaxy \citep[e.g.][]{Marnoch23} still would benefit from an \hi\ absorption line identification. However, given the evolution of the field, there are new pathways that an \hi\ absorption detection in an FRB spectrum may illuminate.}

\subsection{Constraining \tspin\ with FRBs}

\red{Given that host galaxies are now regularly localised, it is possible to measure the \hi\ column density \nhi\ along an FRB's line of sight through direct mapping of the host's \hi\ emission. By combining such a measurement with an absorption constraint, it is possible to infer the spin temperature \tspin\ of that \hi\ reservoir through the expression}

\begin{equation}
    N_\text{HI} = 1.823 \times 10^{18}~ T_\text{spin}\int \tau(\nu)~d\nu \ \  \text{cm}^{-2}\ .
\end{equation}

\red{This quantity directly characterises the thermal state of the gas through its relation to the relative populations of the ground and excited hyperfine levels of atomic hydrogen. Notably, it is empirically very difficult to measure \citep{Allison2021}; simultaneously constraining both the \hi\ column density and the integrated absorption strength is typically only possible when a background radio continuum source (AGN) sufficiently illuminates the gas within its own disk \citep[e.g.][]{Yoon2025}, or, in rare geometric configurations, lies directly behind a galaxy \citep[e.g.][]{Curran2010,Kanekar2014}. In these cases, foreground Lyman-$\alpha$ absorption of the backlight provides the column density estimate, though there have been a few cases where \hi\ emission from a foreground galaxy has been detectable \citep{Reeves2016,Gupta2018}.}

\red{FRBs, however, offer an alternative path; any burst whose bandwidth encompasses the redshifted \hi\ line inherently provides an absorption constraint, and thus simply observing the host galaxy's \hi\ emission immediately provides a limit on \tspin. To demonstrate this, we extend our 3$\sigma$ opacity limit for FRB 20211127I to a corresponding \tspin\ limit. We multiply the opacity by our assumed line width of 50 \kms\ to estimate an integrated optical depth upper limit of roughly 26 \kms. Then, again assuming an optically thin medium, we use the brightness temperature of the \hi\ emission line (see Figure \ref{fig:spectra}) to determine \nhi, which we find to be 1.43\tenpow{21} cm$^{-2}$. Combining these values together results in a 3$\sigma$ lower limit on \tspin\ $\approx$ 30 K.}

\red{Unsurprisingly, this limit yields little of physical value. Derived \tspin\ values from direct \hi\ emission-absorption systems are in the hundreds of K \citep{Reeves2016,Gupta2018}, where interesting conclusions can be drawn about the dynamics between various phases in the neutral medium. Milky Way absorption finds typical lower limits for atomic cloud temperatures around 80 K \citep{Murray2018}, and thus our lower limit of 30 K provides essentially zero constraining power. A positive absorption detection, however, would yield a far more meaningful constraint.}

\red{Of course, such inference is subject to several assumptions. Chief among these are the difference in scales probed by a pencil-beam absorption profile versus a beam-averaged emission profile, and the inherent uncertainty in physical interpretation arising from ignorance of the true cloud structure along the line of sight. However, these drawbacks are largely present regardless of how one measures \tspin\ --- and FRBs in fact sidestep the additional uncertainty introduced by the unknown covering factor of the absorbing gas. The very few studies that estimate \nhi\ from an \hi\ emission map \citep{Reeves2016,Gupta2018} stress that the measured \tspin\ value is a column-density-weighted harmonic-mean temperature, describing the broader thermal state of the ISM in the region around the absorber rather than any individual cloud along the line of sight.}

\red{Currently, such analysis would be limited to low redshift ($z<0.1$) FRBs. Beyond this distance, detecting resolved \hi\ emission in host galaxies becomes unreliable, owing to both the inherent weakness of the \hi\ transition and the prevalence of terrestrial RFI at the frequencies to which the line is redshifted, out to $z\sim0.2$. This redshift restriction is not necessarily a disadvantage, however. Most \tspin\ measurements in the literature using Ly-$\alpha$ absorption of backlights to estimate the foreground \nhi are confined to $z\gtrsim1$, as the Ly-$\alpha$ line rests in the ultraviolet and thus must be redshifted into the optical before becoming accessible to ground-based observatories. Thus, \hi\ absorption in FRBs allows one to probe a complementary and largely unexplored window at low redshift.}

\subsection{Probing the Local Environment Around FRBs}

\red{Though significant progress has been made towards determining the progenitors from which FRBs arise \citep{Niu2022,Bruni2025}, there remain several open questions surrounding their formation channels and birth environments. Detecting \hi\ absorption at the redshift of the host galaxy may inform on the nature of the cold gas in and around the circumburst medium itself, from which inference can be made about the progenitor.}

\red{FRBs are known to preferentially arise in star-forming galaxies \citep{Gordon2023,Sharma2024}, and at least a considerable fraction ($\sim20-46\%$) are specifically associated with spiral arm structure within their hosts \citep{Gordon2025}. As such, it is largely expected that young magnetars formed through core collapse supernovae constitute a large fraction of FRB progenitors. Such objects usually lie within an expanding shell of ionised material \citep{Margalit2018}, shockwave-heated to temperatures likely prohibitive to the presence of any significant absorbing atomic gas in the immediate surroundings. However, these structures are typically embedded within larger star-forming regions --- such as in the spiral arms --- that are almost certain to contain sufficient gas for meaningful absorption to occur \citep{Bochenek2021}. Conversely, delayed channel progenitors (i.e. compact merger remnants) are not expected to lie within gaseous structure at all \citep{Berger2014}}.

\red{Discriminating against local and ISM \hi\ absorption is non-trivial, but there are some avenues through which this could be achieved. Firstly, an absorption signal in a hyperactively repeating FRB would allow for any short term evolution in optical depth or line width to be tracked, whose presence would directly point to local structure variation very near the progenitor. Such evolution has been widely observed in the rotation measures \citep[e.g.][]{Michilli2018,Uttarkar26} and dispersion measures \citep[e.g.][]{Kumar23,Pandhi2026} of repeating FRBs, and thus it is plausible to expect similar behaviour in the properties of an \hi\ absorption line arising from nearby gas.} 

\red{Another approach may be possible in the case that a resolved absorption feature is present in a local Universe FRB, as one could then compare its velocity structure with \hi\ emission arising from the same on-sky region.  An absorption line whose central velocity is significantly offset from the centre of the emission would indicate that the absorbing gas is kinematically distinct from the bulk ISM at that location, lending support to an origin in material local to the progenitor rather than the general ISM column.}

\subsection{FRB Distance Within Host}

A significant issue facing those who aim to use FRBs as cosmological probes is the unknown contribution of the host galaxy on the dispersion measure (DM) of FRBs \citep{James2022,Connor2025}. Disentangling this factor from the cosmological contribution, which is sensitive to several key cosmological parameters, is an ongoing thorn in the side of modellers. Part of this issue arises from the unknowable projected distance an FRB lies within its host; the dispersion of a burst's signal may be significantly larger if it lies on the far side of a galaxy rather than the near side. 

\red{At the most basic level, a correlation between \hi\ absorption strength and excess DM across a sample of localised FRBs would provide a direct and empirical basis for host DM subtractions. More directly, the \hi\ emission-absorption comparison introduced in the previous subsection is also relevant here; in cases where an absorption profile spans only a subset of a corresponding emission line width, the missing velocity components indicate gas lying on the far side of the FRB, constraining its depth within the host. In sufficiently resolved systems, this could be combined with kinematic disk modelling \citep[e.g. with \texttt{3DBarolo},][]{Barolo} to yield a probabilistic line-of-sight placement.}  

\red{Alternatively, if one instead assumes a value for \tspin, as is commonplace, an \hi\ column density measurement derived via emission can be compared with an inferred column density estimate as a function of \tspin. If the ratio of the inferred to measured \nhi\ approaches zero (unity), the FRB likely lies on the near (far) side of the bulk ISM. While this evidently requires an assumption about the thermal conditions of the intervening gas, it may nonetheless allow discrimination between highly disparate origins, e.g. a foreground globular cluster versus a source embedded within the gaseous disk. In the case of FRB~20211127I, the ratio would have an upper limit of} 

\[
    \frac{N_\text{HI,inf}}{N_\text{HI,meas}} < 0.033~T_\text{spin}
\]

\noindent\red{which exceeds unity for reasonable \tspin\ values in the hundreds of K \citep{Allison2021}, and thus offers no meaningful geometric constraint as expected.}

Regardless of inference method, comparing some estimated kinematic depth with the observed DM would provide stronger host galaxy constraints than are currently possible, particularly given the lack of correlation between excess DM and other FRB properties \citep{Scott2025,Mas-Ribas2025}. Furthermore, this depth may also provide a method to infer the impact of host galaxy scattering. By definitively placing the FRB relative to the host's \hi\ reservoir, one could determine whether observed pulse broadening is \red{likely to be} dominated by the dense, turbulent environment immediately surrounding the progenitor or the integrated column of the galactic disk. \red{Disentangling these contributions would provide useful constraints on the scattering-DM relation, which remains poorly understood at extragalactic distances.}

\section{Conclusion}\label{sec:conclusion}

\red{We revisit the prospect of detecting \hi\ absorption features in the spectra of localised FRBs, and discuss their scientific utility in the modern FRB field. We conduct a preliminary search for \hi\ absorption in the spectrum of FRB~20211127I, a bright ASKAP-localised burst whose host harbours a large \hi\ reservoir along the FRB's line of sight. The host redshift places the \hi\ line at the boundary of a scintle in the FRB's spectrum, making this an instructive test case. Detecting no feature, we place a 3$\sigma$ upper limit on the opacity of a theoretical absorption line of 0.51. We further combine this constraint with \hi\ column density estimates from a 3 hr $L$-band observation with MeerKAT to estimate a lower limit on the surrounding gas \tspin\ of 30 K.} While this non-detection unsurprisingly offers little constraining power, we discuss the possibility of detecting absorption in the signals of FRBs given the capabilities and strategies of modern observatories.

We first calculate the fraction of FRBs detected by ASKAP, DSA, FAST, and MeerKAT capable of probing opacities below 0.1, finding that the incoherent detection modes of MeerKAT and ASKAP show the greatest proportions at 10$\%$ and 6$\%$. By also factoring in FoV and post-detection processing capabilities, we find that ASKAP, observing at full 36-dish strength, currently offers the greatest chance of detecting \hi\ absorption in a single FRB. This would still require a relatively narrow ($<3$ ms) burst with a fluence greater than 100 Jy ms. However, we reiterate that such FRBs have been previously observed; the $z\sim 0.06$ MeerKAT FRB 20210405I would have probed deep absorption and \tspin\ limits had its voltages been saved to produce a higher resolution spectrum. 

We also consider the opportunity presented by the repeating FRB population, particularly the hyperactive sub-population which repeat thousands of times. Monitoring these objects offers a separate avenue that, through FAST's current sensitivity, \red{makes absorption detection a very realistic prospect through the stacking of many outbursts; we highlight that the hyperactive repeater FRB~20220912A is an ideal candidate for such an investigation.}

\red{Finally, we comment on the prospective science such a detection could offer. A confirmed \hi\ absorption detection in a nearby, localised FRB would open a direct window onto the thermal state of the host ISM through a measurement of \tspin.} Furthermore, combining FRB \hi\ absorption with host galaxy \hi\ emission maps may provide a means to disentangle the host galaxy contribution to an FRB's dispersion measure from its cosmological component, directly aiding the use of FRBs as cosmological probes. 

\section*{Acknowledgements}
We thank Ben Stappers,  Laura Driessen, and Adam Deller for providing data and useful discussions that impacted the analysis of this work. We also thank the reviewer for helpful comments. H.R.\ is supported by an Australian Government Research Training Program (RTP) Scholarship. M.G.\ and C.W.J.\ acknowledge support by the Australian Government through the Australian Research Council Discovery Projects funding scheme (project DP210102103). A.B.\ acknowledges support through project CORTEX (NWA.1160.18.316) of the research programme NWA-ORC which is financed by the Dutch Research Council (NWO). M.G.\ also acknowledges support through UK STFC Grant ST/Y001117/1. M.G. acknowledges support from the Inter-University Institute for Data Intensive Astronomy (IDIA). IDIA is a partnership of the University of Cape Town, the University of Pretoria and the University of the Western Cape. For the purpose of open access, the author has applied a Creative Commons Attribution (CC BY) licence to any Author Accepted Manuscript version arising from this submission. A.B. acknowledges support through the project CORTEX (NWA.1160.18.316) of the research programme NWA-ORC which is financed by the Dutch Research Council (NWO).

This scientific work uses data obtained from Inyarrimanha Ilgari Bundara, the CSIRO Murchison Radio-astronomy Observatory. We acknowledge the Wajarri Yamaji People as the Traditional Owners and native title holders of the Observatory site. CSIRO's ASKAP radio telescope is part of the Australia Telescope National Facility (https://ror.org/05qajvd42). Operation of ASKAP is funded by the Australian Government with support from the National Collaborative Research Infrastructure Strategy. ASKAP uses the resources of the Pawsey Supercomputing Research Centre. Establishment of ASKAP, Inyarrimanha Ilgari Bundara, the CSIRO Murchison Radio-astronomy Observatory, and the Pawsey Supercomputing Research Centre are initiatives of the Australian Government, with support from the Government of Western Australia and the Science and Industry Endowment Fund. We also thank the MRO site staff. The MeerKAT telescope is operated by the South African Radio Astronomy Observatory, which is a facility of the National Research Foundation, an agency of the Department of Science, Technology and Innovation.

\printbibliography

\end{document}